\documentclass[12pt]{article}

\def\ii{\'{\i}}

\usepackage{epsfig}

\def\arrowright{\smash{\mathop{\longrightarrow}\limits_{\raise3pt\hbox{$\eta \to 0$}}}}

\def\infinito{\smash{\mathop{\longrightarrow}\limits_{\raise3pt\hbox{$\eta \to \infty$}}}}

\def\Ninfinito{\smash{\mathop{\longrightarrow}\limits_{\raise3pt\hbox{$N_A \to \infty$}}}}

\begin{document}

\title{Multiplicity fluctuations in hadron-hadron and nucleus-nucleus collisions and percolation of strings}%
\author{Pedro Brogueira\footnote{Department of Physics, Instituto Superior T\'{e}cnico, 1049-001 Lisboa, Portugal} and  
Jorge Dias de Deus$^*,$\footnote{CENTRA, Centro Multidisciplinar de Astrof\ii sica}}
\maketitle

\begin{abstract}
We argue that recent NA49 results on multiparticle distributions and fluctuations, as a function of the number of participant nucleons, suggest that percolation plays an important role in particle production at high densities. 
\end{abstract}

Recentely, the NA49 collaboration has presented results, from the CERN/SPS at 158 A GeV, on multiplicity fluctuations or, to be more precise, on $V(n)/<n>$, 

\begin{equation}
V(n)/<n> \equiv \frac{<n^2>-<n>^2}{<n>},
\end{equation}
as a function of the number $N_{part.}$ of participant nucleons, from $pp$ to $PbPb$ collisions [1].

These data are very interesting for several reasons:

1) They show evidence for universal behaviour: the experimental points in the plot $V(n)/<n>$ versus $N_{part.}$ fall into a unique curve (see Fig.1).

\begin{figure}[t]
\begin{center}
\includegraphics[width=11cm]{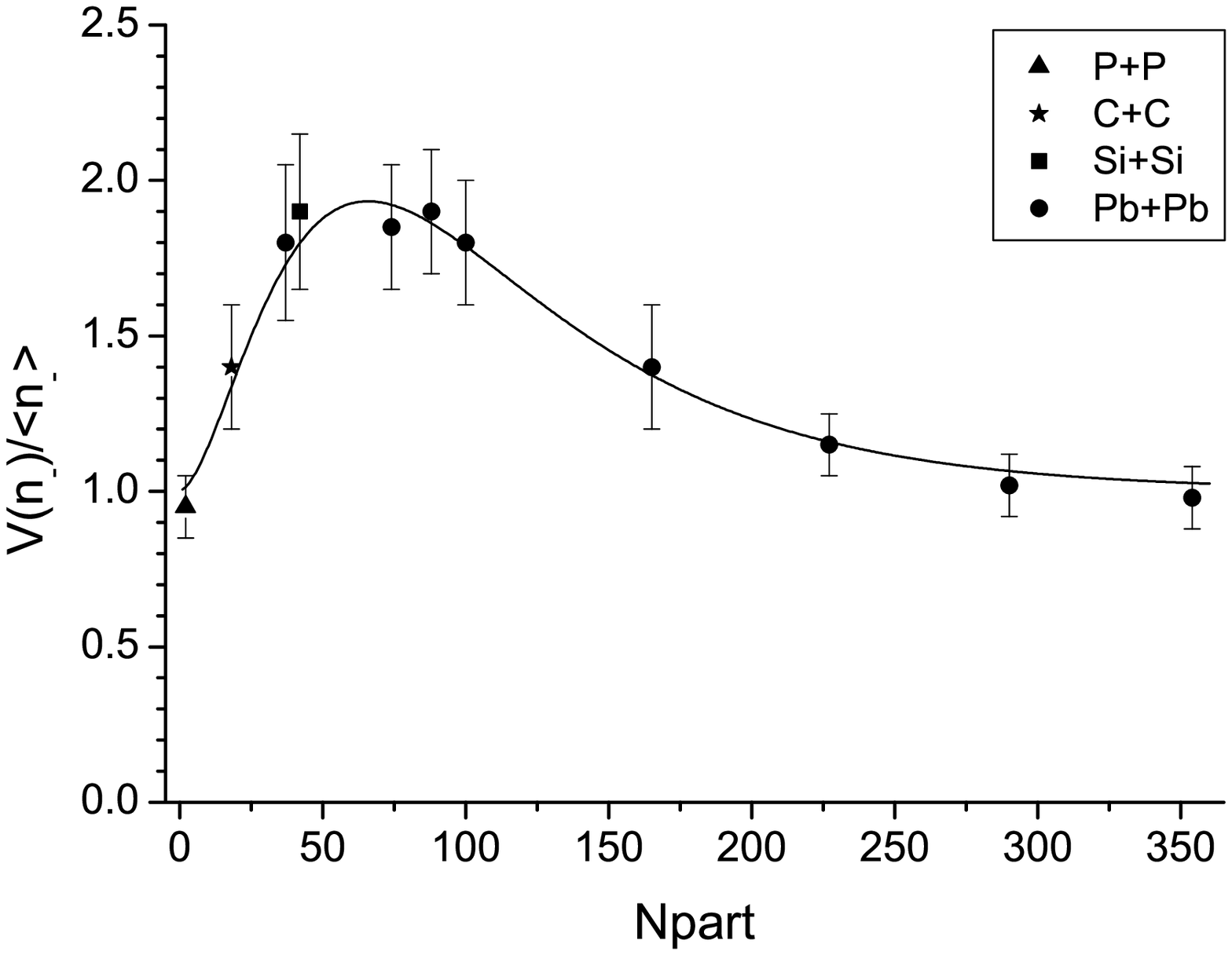}      
\end{center}
\end{figure}

2) The physics in the small $N_{part.}$ limit $(pp,N_{part.}\to 2)$ and in the large $N_{part.}$ limit $(PbPb, N_{part.}\to 2A_{PbPb})$ seems to be quite the same, as in both cases the quantity (1) approaches 1. The fluctuations are larger in the intermediate $N_{part.}$ region (see Fig.1).

3) The (negative) particle distribution, in the low density and in the high density limits, is in fact a Poisson distribution (see Fig.2), the distribution being wider than Poisson in the intermediate $N_{part.}$ region.

In the framework of the string model with percolation [2], these results are quite natural. On one hand, percolation is a universal geometrical phenomenon, the properties depending essentially on the space dimension (dimension 2, impact parameter plane, in our case), and being controlled by the transverse density variable $\eta$,

\begin{equation}
\eta \equiv \left( {r\over R}\right)^2 \bar N_S \ ,
\end{equation}
where $r$ is the transverse radius of the string $(r\simeq 0.2 fm)$, $R$ the radius of the interaction area, and $\bar N_S$ the average number of strings. The quantity $(R/r)^2$ is nothing but the interaction area in units of the string transverse area. As $R$ and $\bar N_S$ are functions of the number $N_{part.}$ of participants, $N_{part.}$, similarly to $\eta$, becomes, at a given energy,a universal variable.

On the other hand, in percolation [3], what matters is the fluctuation in the size of the clusters of strings: one starts, at low density (small $N_{part.}$), from a situation where strings are isolated, at intermediate density one finds clusters of different sizes, and one ends up, at high density, above the percolation threshold, with a single large cluster. In both, low and high, density limits, fluctuations in cluster size vanish (see Fig.3). In the simplest string model the particle distribution is Poisson (as observed in $e^+e^-$ and $pp$ at low energy) and $V/<n>\to 1$ in both, low and high, density limits (see Figs.1 and 2).

\includegraphics[width=11cm]{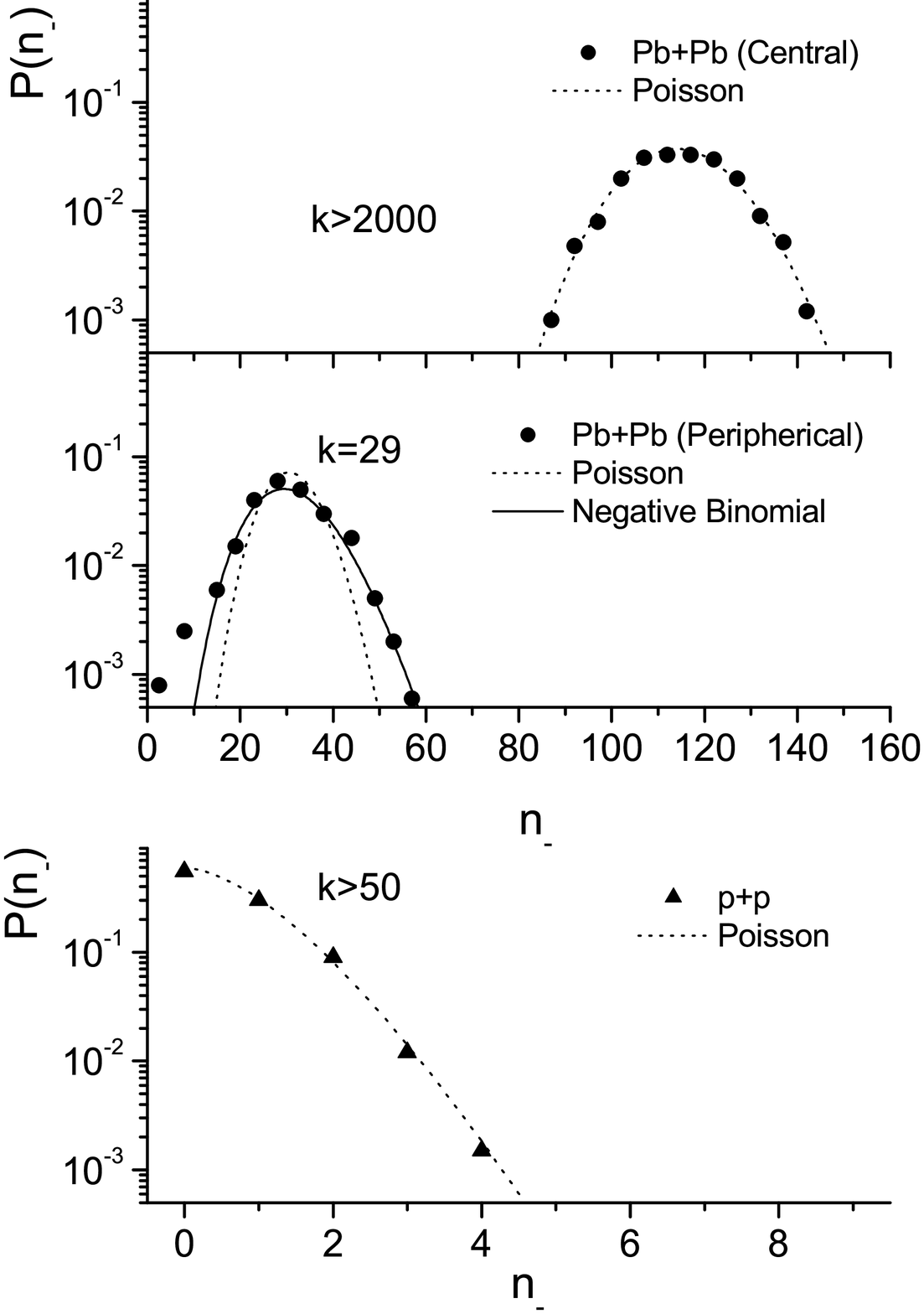}      

Let us try to be more specific. In hadron-hadron and nucleus-nucleus collisions, during the collision strings are produced along the collision axis, and these strings may overlap and form clusters of different sizes. In the spirit of percolation theory, we shall assume that fluctuations in the number $N$ of strings per cluster dominate over fluctuations in the number $N_C$ of clusters.

Generalizing [4], we write for the multiplicity distributions, for a given value of $\eta$ or $N_{part.}$,

\begin{equation}
P(n)=\int dN_C \varphi (N) \int \Pi_{i=1}^{N_C} d N_i d \mu_i W(N_i)P_{N_i}(\mu_i) \delta \left( n-\sum_{i=1}^{N_C} \mu_i \right) \ ,
\end {equation}
where $\varphi (N_C)$ is the probability of having $N_C$ clusters, $W(N_i)$ the probability of having a cluster of $N_i$ strings, $P_{N_i} (\mu_i)$ the probability of a cluster of $N_i$ strings emitting $\mu_i$ particles. We shall further write

\begin{equation}
P_N (\mu) = \int \Pi_{i=1}^{N} P_1 (n_j) \delta \left( \mu -\sum_{j=1}^{N} n_j \right) \ ,
\end{equation}
where $P_1 (n)$ is the single string distribution, assumed to be Poisson.

Note that in (3) we treat the cluster as independent, and in (4) we treat the strings themselves as independent (this is to be corrected later by including string fusion).

As mentioned above, we neglect fluctuations  in the number of clusters. This means that we approximate $\varphi (N_C)$ by a $\delta$-function, such that (3) becomes 
$$
\hskip 2 true cm P(n)= \int \Pi_{i=1}^{\bar N_C} dN_i d\mu_i W(N_i) P_{N_i} (\mu_i) \delta \left( n-\sum_{i=1}^{\bar N_C} \mu_i \right) \ ,\hskip 1.1 true cm (3')
$$
where $\bar N_C$ is the average number of clusters, for a given density $\eta$, or number of participant nucleons $N_{part.}$.

From ($3'$) and (4) we obtain,

\begin{equation}
<n> = \bar N_C <N> \bar n \ ,
\end{equation}
and 
\begin{equation}
<n^2>-<n>^2 = \bar N_C [(<N^2>-<N>^2)\bar n^2 + <N> (\bar n^2 - \bar n^2)] \ ,
\end{equation}
where $\bar n$ is the single string particle multiplicity, and, finally, for the (variance)/(multiplicity) ratio,

\begin{equation}
{V(n) \over <n>} = \bar n {V(N)\over <N>} + 1 \ ,
\end{equation}
where
\begin{equation}
{V(N)\over <N>} \equiv {<N^2> - <N>^2 \over <N>} \ .
\end{equation}

The previous percolation argument for the fluctuations in the size of the clusters (see Fig.3) requires:
\begin{equation}
{V(N) \over <N>} \ \arrowright 0 \ \ \ {\rm and}\ \ \ {V(N) \over <N>} \ \ \infinito  0 \ \ .
\end{equation} 

\includegraphics[width=6cm,angle=270]{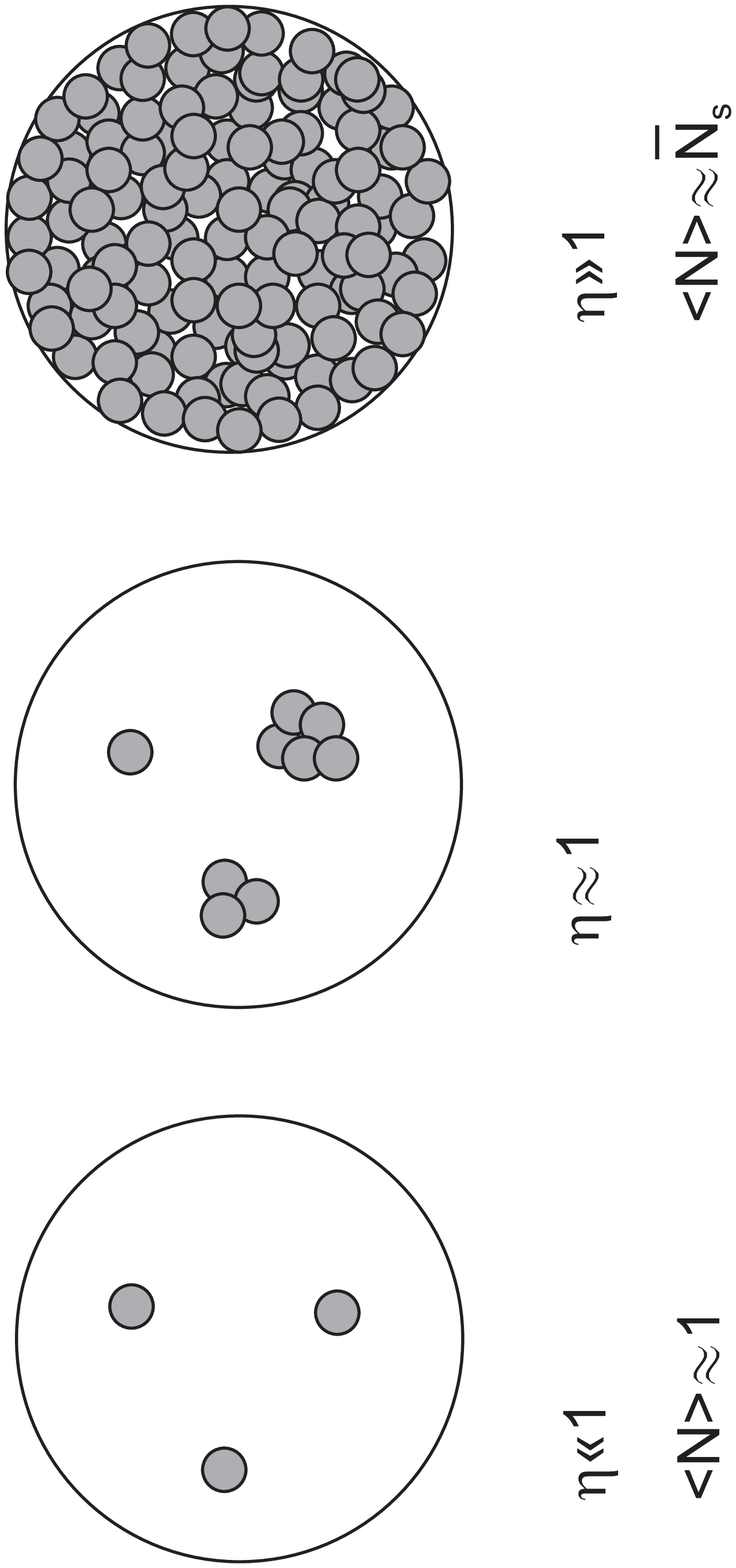}

In percolation there are two sum rules to be satisfied. If $N$ stands for the number of strings in a cluster, $N_C$ the number of clusters and $N_S$ the number of strings, we have:
\begin{equation}
\bar N_C <N> = \bar N_S \ ,
\end{equation}
and
\begin{equation}
\bar N_C <A> = ({R\over r})^2 (1-e^{-\eta}) \ ,
\end{equation}
where $<A>$ is the average area occupied by a cluster (see, for instance, [3,5]). In percolation $<A>$ increases with $\eta$, approaching the full area of interaction in the $\eta \to \infty$ limit.

In the case of percolation we thus write for $<A>$, [5],
\begin{equation}
<A> = f(\eta) [({R\over r})^2 (1-e^{-\eta})-1] +1\ ,
\end{equation}
where $f(\eta)$ is a percolation function, such that $f(\eta ) \to 0$, as $\eta \to 0$, and $<A> \to 1$, as expected for isolated strings, and $f(\eta ) \to 1$, as $\eta \to \infty$, and $<A>\to ({R\over r})^2$ as expected in percolation. For $f(\eta )$ we have chosen
\begin{equation}
f(\eta ) = (1+e^{-(\eta -\eta_{c})/a})^{-1} \ ,
\end{equation}
with $a=0.85$ and $\eta_c = 1.15$ [5].

From (10), (11) and (12) we obtain
\begin{equation}
<N> = {\eta \over 1-e^{-\eta}} (f(\eta ) [({R\over r})^2 (1-e^{-\eta})-1]+1) \ ,
\end{equation}
with $<N> \to 1$ as $\eta \to 0$, and $<N> \to \bar N_S$, as $\eta \to \infty$ (see Fig.3).

Regarding the cluster string variance, $<N^2> - <N>^2$, we have to satisfy the general constraint,
\begin{equation}
<N^2> - <N>^2 \geq 0 \ ,
\end{equation}
and in addition we imposed the constraints,
\begin{equation}
<N^2>-<N>^2 \arrowright 0 \ ,
\end{equation}
and 
\begin{equation}
<N^2> - <N>^2 \infinito 0 \ .
\end{equation}
We thus write for the variance,
\begin{equation}
V(N)\equiv <N^2> -<N>^2 = \left[ {1\over b} {1-(1+b\eta )e^{-b\eta}\over e^{b\eta} -1} \right] <N>^2 \ ,
\end{equation}
where $b > 0$ is an adjustable parameter (fixed at the value $b=1.65$). Note that (18) satisfies (15), (16) and (17). 

Before making a comparison between our string percolation model and NA49 data, there are two questions to be addressed:

i) {\it $F(\eta)$ factor due to random colour summation}

When strings fuse in a cluster the effective colour charge is not just the sum of the colour charges of the individual strings [6]. In pratice, the effective number $N$ of strings is reduced, [7],
\begin{equation}
N \longrightarrow \sqrt{<A>\over <N>} N\longrightarrow F (\eta) N \ ,
\end{equation}
where (see (10) and (11)),
\begin{equation}
F(\eta) = \sqrt{1-e^{-\eta}\over \eta} \ ,
\end{equation}
such that, instead of (5) and (7), have
$$
\hskip 3 true cm <n> = F(\eta) \bar N_C <N> \bar n = F(\eta) \bar N_S \bar n \ , \hskip 2 true cm (5')
$$
and 
$$
\hskip 4.3 true cm {V(n)\over <n>} = F(\eta) \bar n {V(N)\over <N>} + 1 \ . \hskip 3 true cm (7')
$$
Note that the $F(\eta)$ correction is more important for $<n>$, ($5'$), then for Eq. ($7'$). If the single particle distribution is Poisson with average multiplicity $\bar n$, the average cluster has also a Poisson distribution with multiplicity $F(\eta) <N> \bar n$.

ii) {\it The relation between $\eta$ and $N_{part.}$}

In the definition of $\eta$, (2), what appears is not $N_{part.}$ but rather the average number $\bar N_S$ of strings and the radius $R$ of interaction. Making use of simple nuclear physics and multiple scattering arguments, one has [8]
\begin{equation}
R \simeq R_p N_A^{1/3} \ ,
\end{equation}
and
\begin{equation}
\bar N_S \simeq \bar N_S^p N_A^{4/3} \ , 
\end{equation}
where $R_p$ is the nucleon radius $(\simeq 1fm)$, $\bar N_S^p$ is the (energy dependent) number of strings in $pp$ collisions, at the same energy, and $N_A$ is given by 
\begin{equation}
N_A = {N_{part.}\over 2} \ .
\end{equation}
We would like to mention that Eqs.(21) and (22) are not rigorous: in (21) geometrical factors are not taken into account, in (22) no distinction is made between valence strings and sea strings, [8].

From (2), (21), (22) and (23) we obtain for the relation between $\eta$ and $N_{part.}$,
\begin{equation}
\eta = \left( {r\over R_1}\right)^2 \bar N_s^p N_A^{2/3}
\end{equation}

We shall now present our results:

\medskip

{\bf 1- $\bf V(n)/<n>$}
\smallskip

In Fig.1 we show our curve, Eq.($7'$) with (14), (18) and (24), in comparison with NA49 data. The obtained values for $\bar n(\bar n= 0.12)$ and $N_s^p (N_s^p = 4.5)$ are consistent with the additional constraint $<n>_p\simeq 0.52$ (as seen in Fig.2). Note that the behaviour of our curve is essentially determined by the behaviour of $\left( <N^2> -<N>^2 \right) /<N>$.

In Fig.2 we show fits of the multiplicity distributions for different values of $N_{part.}$, with Negative Binominals. Reasonable fits are obtained with values for the NB parameter: $k=\infty$, Poisson, at low and high density, and $k=29$, for intermediate density.

\medskip

{\bf 2- $\bf <n>_{N_A}$ \ and \ $\bf <n>_p$}
\smallskip

We can relate $<n>_{N_A}$ to $<n>_p$ by making use of  (5') and (24):
\begin{equation}
<n>_{N_A} = {F(\eta_{N_A})\over F(\eta_p)} <n>_p N_A^{4/3}
\end{equation}
we first note that (25) satisfies saturation when $N_A \to \infty$:
\begin{equation}
{1\over N_A} <n>_{N_A} \Ninfinito const.
\end{equation}
Relation (25) is valid for high energy. At low energy -- $\sqrt{s}\simeq 20$GeV is the energy at SPS -- the presence of valence quarks cannot be ignored. We take them into account by writing, instead of (25),
$$
\hskip 1.3 true cm <n>_{N_A} = {F(y_{N_A})\over F(n_p)} <n>_p N_A^{4/3} \left[ 1-c\left( 1-1/N_A^{1/3}\right) \right] \ , \hskip 1.3 true cm (25')
$$
where $c$ is a parameter decreasing with energy, $1\geq c\geq 0$, measuring the relative contribution of valence quarks to multiplicity,
\begin{equation}
c\equiv {<n>^V_p \over <n>_p} \ ,
\end{equation}
and for $c$ was taken the value $c=0.59$. In Fig.4 we show ($25'$) in comparison with NA49 data. 

\includegraphics[width=11cm]{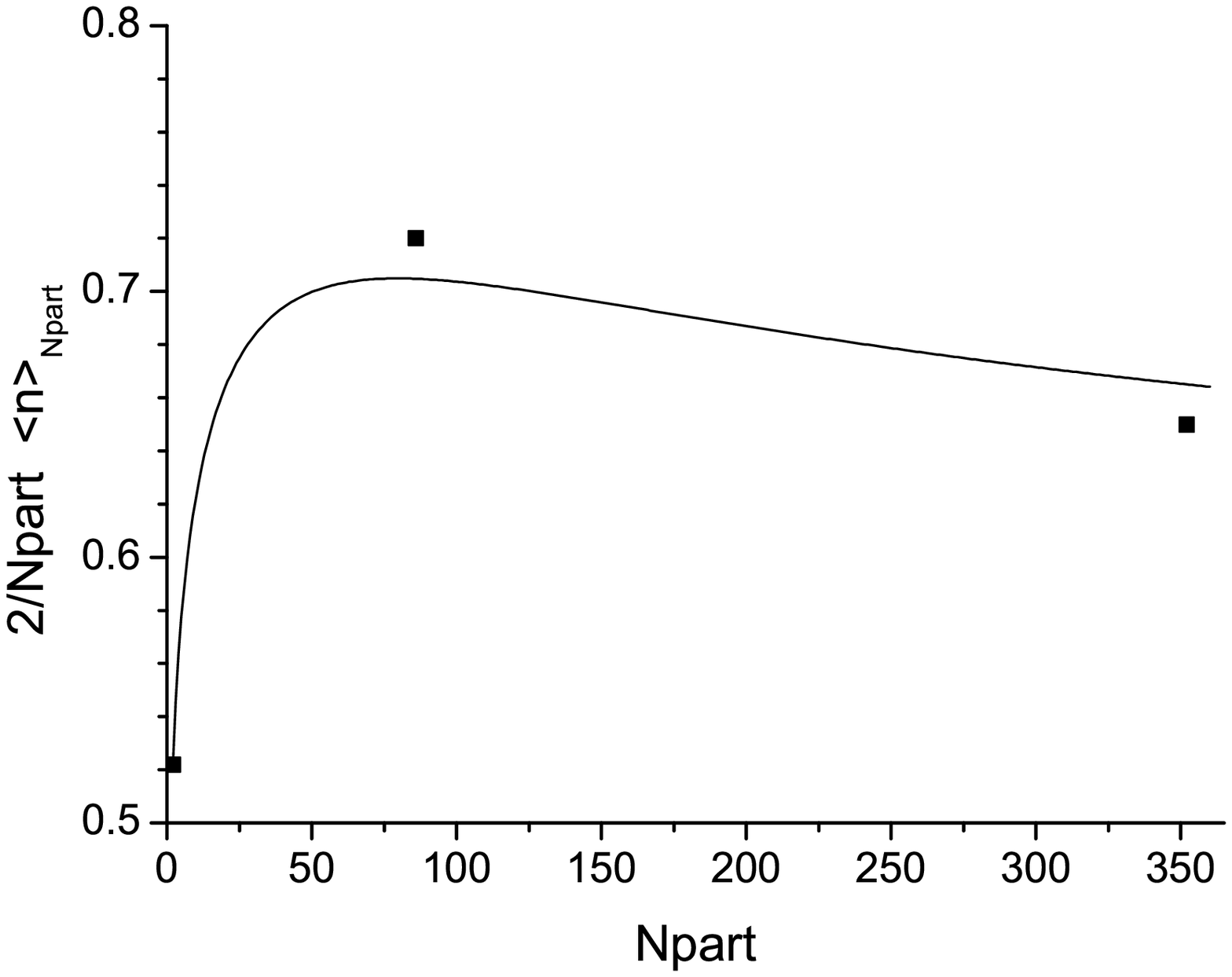}

{\bf 3- $\bf P_T$ fluctuations}
\smallskip

It has been argued [9] that a convinient quantity to measure genuine $P_T$ correlations -- above random correlations -- is $\phi_{P_T}$
\begin{equation}
\Phi_{P_T} \equiv \sqrt{{<Z^2>\over <n>}}- \sqrt{\bar z^2} \ ,
\end{equation}
where, for a given event, $Z\equiv \sum_{i=1}^n \left(P_{T_i} - \bar P_T\right)$,
$z\equiv P_T - \bar P_T , \bar P_T$ meaning averaging over the single particle inclusive distribution. NA49 has presented results for $\phi_{P_T}$ [10], in the same conditions of the multiplicity experiment.

A crude percolation model for $P_T$ fluctuations, developed in [5], gives
\begin{equation}
\phi_{P_T} = \bar P_T \left[ \sqrt{\bar n {V(N)\over <N>} + \ell} - \sqrt{\ell}\right] \ ,
\end{equation}
where $\ell = \left( \overline{P_T^2} - {\overline P}_T^{^2} \right) /{\overline P}_T^{^2}$. In  Fig.5 we compare (28) with data. Note that in (29) $\phi_{P_T} \to 0$ at small and high densities, as seems to be the case in data.

\includegraphics[width=11cm]{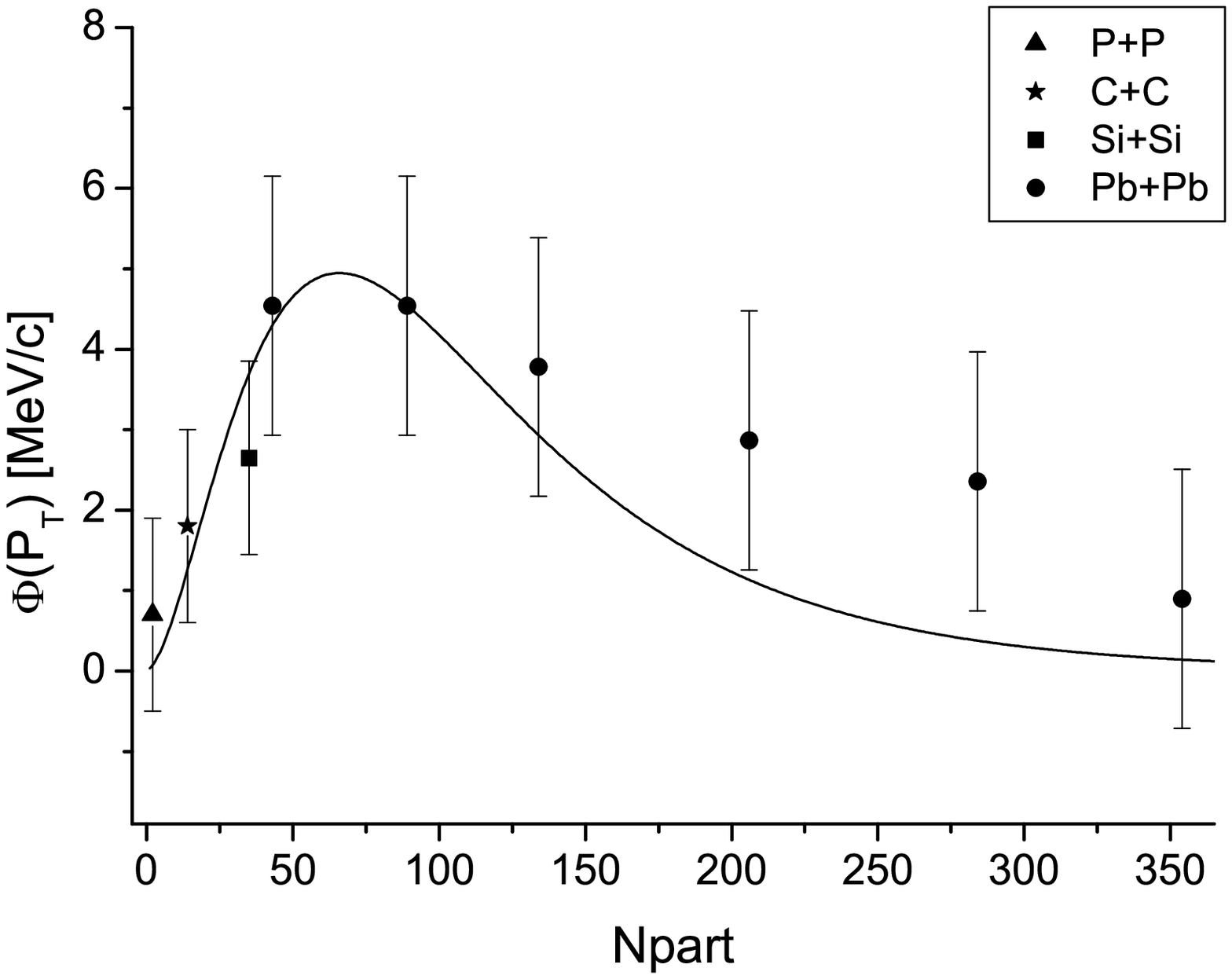}

In a recent paper [11] an attempt was made to relate $\phi_{P_T}$ to $V(n)/<n>$. However a direct proportionality, as proposed in [11], does not seem to satisfy the low and high density limits. We would like to  mention that the use of percolation to study $P_T$ fluctuations was initiated with [12]. 
\medskip


{\bf 4- Rapidity interval and centre of mass energy dependence of $\bf <n>$ and $\bf k$}
\smallskip

The average multiplicity $<n>$, ($5'$),
$$
<n> = F(\eta ) \bar N_C <N> \bar n \ ,
$$
and the Negative Binomial Parameter $k$,
\begin{equation}
1/k \equiv {<n^> - <n>^2 \over <n>^2} - {1\over <n>} \ ,
\end{equation}
or, from ($7'$),
\begin{equation}
k=\bar N_C {<N>^2 \over <N^2> -<N>^2}\ ,
\end{equation}
are both proportional to $\bar N_C$, the average number of clusters. If we assume that the clusters of strings are approximately uniformely distributed in the central rapidity region -- this is the central rapidity plateau idea -- then $\bar N_C$ is, for $\Delta y>0$, a linear function of $\Delta y$ (but approching 1 as $\Delta y \to 0$, i.e., $<N>\to 1$). The multiplicity is also a linear function of $\Delta y$ (and approaching zero as $\Delta y \to 0$, because $\bar n$ is also a linear function of $\Delta y$, for small $\Delta y$). The expected $<n>$ and $k$ dependence on $\Delta y$ - linear increase with $\Delta y$ for large $\Delta y$ - agrees with data [13].

On the other hand, $<n>$ is an increasing function of $\sqrt{s}$, because $\bar N_C <N> = \bar N_S$ increases with $\sqrt{s}$ and, at large $\eta$, $F(\eta)$ only decreases as $1/\sqrt{\bar N_S}$. Increasing the energy, at fixed number of participants, means increasing $\eta$. Regarding $k$, $\bar N_C$ and $<N>^2/(<N^2>-<N>^2)$,these are not monoatonic functions of $\eta$. The NB parameter $k$ goes to infinity at small $\eta$ (energy) and at large $\eta$ (energy). One thus expects $k$, for instance in $pp$, to have a minimum at some energy [14].

In conclusion, we find that the recent NA49 results, regarding the multiplicity distribution dependence on the number of participant nucleons are quite consistent with the impact parameter, percolation description of hadron-hadron and nucleus-nucleus collisions at high energies and high densities.

We would like to thank Elena Ferreiro, Carlos Pajares and Roberto Ugoccioni for many discussions, and to thank P. Seybot and M. Rybcz\'ynski for information on NA49 data. This work has been done under the contract POCTI/36291/FIS/2000, Portugal.

\vfill \eject

\textbf{Figure Captions:}

\begin{description}
\item[Fig.1] Variance over average multiplicity, for negative particle production, as a function of the number of participants. The curve is from ($7'$) with (14), (18) and (24). Data are from NA49 [1].
\item[Fig.2] Multiplicity Distributions, $P(n_-)$, as a function of $n_-$. The curves are Poisson (dashed lines) and Negative Binomial (full line).
\item[Fig.3] Impact parameter percolation. For small densities $(\eta \ll 1)$ and for large densities $(\eta \gg 1)$ there are no strong fluctuations in the number $N$ of strings per cluster. For intermediate densities $(\eta \simeq 1)$ $N$-fluctuations are large.
\item[Fig.4] The average multiplicity divided by $1/2 N_{part.}$ as a function of $N_{part.}$. The curve corresponds to Eq.($25'$). Multiplicities were calculated from the distributions of Fig.2.
\item[Fig.5] $\phi \left( P_T\right)$ as a function of $N_{part.}$. The curve corresponds to Eq.(29). Data are from [10].
\end{description}

\vfill \eject

\textbf{References:}

\begin{enumerate}
\item M. Ga\v zdzicki et al., Report from NA49, nucl-ex, 0403023 (2004).
\item N. Armesto, M.A. Braun, E.G. Ferreiro and C. Pajares, Phys. Rev. Lett. 77, 3736 (1996); M. Nardi and H. Satz, Phys. Lett. B442, 14 (1998); A. Rodrigues, R. Ugoccioni and J. Dias de Deus, Phys. Lett. B458 (1999) 402.
\item D. Stauffer, Phys. Rep. 54, 2 (1979); D. Stauffer and A. Aharony, Introduction to Percolation theory, Taylor and Francis (1992).
\item J. Dias de Deus, F.G. Ferreiro, C. Pajares and Ugoccioni, hep-ph/0304068; C. Pajares, Acta Phys. Pol. B35, 153 (2004).
\item J. Dias de Deus and A. Rodrigues, hep-ph/0308011 (2003).
\item T.S. Biro, H.B. Nielsen and J. Knoll, Nucl. Phys. B245, 449 (1984).
\item M.A. Braun, F. Del Moral and C. Pajares, Phys. Rev. C65, 02490 (2002).
\item J. Dias de Deus and R. Ugoccioni, Phys. Lett. B491, 253 (2000), Phys. Lett. B494, 53 (2000).
\item M. Ga\v zdzicki and St. Mr\'owczy\'nski, Z. Phys. C52, 127 (1992).
\item H. Appelshauser et al., NA49 Collaboration, Phys. Lett. B459, 679 (1999); T. Anticic et al, NA49 collaboration, hep-ex/0311009 (2003).
\item St. Mr\'owczy\'nski, M. Rybczy\'nski and Z. Wlodarczyr, nucl-th/0407012 (2004).
\item E.G. Ferreiro, F. Del Moral and C. Pajares, Phys. Rev. C69, 034901 (2004).
\item L. Van hamme, UA5 collaboration, in 6th Topical Workshop on Proton-Antiproton collider Physics, pg 491, Ed. K. Eggert, H. Faissner and E. Radermacher (1986), World Scientific.
\item J. Dias de Deus, E.G. Guerreiro, C. Pajares and R. Ugoccioni, Phys. Lett. B581, 156 (2004).
\end{enumerate}

\end{document}